\documentstyle[twocolumn,eqsecnum,aps]{revtex}

\def\half{{\textstyle{1\over 2}}}
\begin{document}
\wideabs{
\title{Nonequilibrium quantum dynamics of phase transitions in an expanding universe}
\author{Ian D. Lawrie}
\address{Department of Physics and Astronomy, The University of Leeds, Leeds LS2 9JT, England}
\date{\today}
\maketitle
\begin{abstract}
I summarize the derivation of a set of Feynman rules appropriate for the perturbative
description of the nonequilibrium dynamics of the symmetry-breaking phase transition in
$\lambda\phi^4$ theory in a Robertson-Walker universe.  The approximation scheme I develop
provides for a treatment of dissipative effects which are essential to an adequate description
of the nonequilibrium state.  It also provides for the emergence from an initially symmetric
state of a final state exhibiting the properties of spontaneous symmetry breaking without the
introduction by hand of any explicit symmetry breaking.
\vskip 10pt

{\it Talk presented at the 5th International Workshop on Thermal
Field Theories and Their Applications, Regensburg, August 1998}

\end{abstract}}

\section{Introduction}
Thermal field theory strongly suggests that the hot, dense matter present in the rapidly
expanding early universe would undergo a variety of phase transitions.  In particular, the
``new'' inflationary scenario of Linde \cite{linde1982} and Albrecht and Steinhardt
\cite{albrecht1982} envisaged a symmetry-breaking phase transition in a grand unified theory
(modifying the earlier, seminal proposal of Guth \cite{guth1981}) which would lead to a
short period of quasi-exponential expansion. Amongst cosmologists, the standard wisdom is that
the classical theory of scalar fields provides an adequate description of the relevant physics,
and that the new inflationary scenario does not yield an acceptable cosmology.  However, a fully
quantum-field-theoretic treatment even of the simplest model of a phase transition in an
expanding universe is rather difficult, and has rarely been attempted.  (A series of papers by
Boyanovsky, Holman, de Vega and others (see Ref. \cite{boyanovsky1998} and references therein)
gives a detailed treatment of a model which can essentially be solved exactly, and I will
later discuss this work in the context of the programme described here.)

The work on which I want to give a progress report has as its goal the solution of the
field equations of semi-classical general relativity, namely
\begin{equation}
{\cal G}_{\mu\nu}(g)=\kappa\langle T_{\mu\nu}\rangle\,,
\label{fieldeqns}
\end{equation}
where ${\cal G}_{\mu\nu}$ is the Einstein curvature tensor (which depends on the metric $g$ and
is treated classically) and $T_{\mu\nu}$ is the stress tensor of an appropriate quantum field
theory.  The simplest version of this problem (which is all that I am competent to discuss)
takes the spacetime geometry to be described by a Robertson-Walker metric, with line element
$ds^2=a^2(t)\left[dt^2-d\bbox{x}^2\right]$.  Here, $t$ is conformal time, $\bbox{x}$ denotes
comoving spatial coordinates and $a(t)$ is the scale factor.  The field theory I will consider
is defined in Minkowski spacetime by the Lagrangian
\begin{equation}
{\cal L}=\frac{1}{2}\partial_{\mu}\Phi\partial^{\mu}\Phi +\frac{1}{2} m_0^2\Phi^2
-\frac{\lambda}{4!}\Phi^4\,,
\end{equation}
where $\Phi$ is a single scalar field, and the bare mass parameter $m_0^2$ is positive, so that
spontaneous symmetry breaking is expected at zero temperature.  In the Robertson-Walker
spacetime, I suppose this field to be conformally coupled to gravity, in which case the scaled
field $\phi(\bbox{x},t)=a(t)\Phi(\bbox{x},t)$ has the action
\begin{equation}
S=\int\,d^4x\left[\frac{1}{2}\partial_\mu\phi\partial^\mu\phi +\frac{1}{2}m^2(t)\phi^2
-\frac{\lambda}{4!}\phi^4\right]\,,
\label{lagrangian}
\end{equation}
which is equivalent to a Minkowski-space field theory with an effective time-dependent mass
$m^2(t)=a^2(t)m_0^2$.

The Hamiltonian of this theory is, of course, explicitly time dependent. I will take expectation
values such as that on the right of (\ref{fieldeqns}) to be
\begin{equation}
\langle {\cal O}(t)\rangle={\rm Tr}\left[e^{-\beta_0H(0)}{\cal O}(t)\right]/
{\rm Tr}\left[e^{-\beta_0H(0)}\right]\,,
\end{equation}
where ${\cal O}(t)$ is a Heisenberg-picture operator whose expectation value we want to know at
time $t$.  I am therefore assuming the existence, at some initial time that I call $t=0$, of an
instantaneous state of thermal equilibrium at a temperature $T_0=1/\beta_0$.  Naturally,
the initial temperature is supposed to be high enough for the system to start in its
symmetric state.  Clearly, such an initial state is somewhat artificial.  I choose it so that I
have, at least, a well-posed problem that I can attempt to solve.

I will admit straight away that I have not solved this problem.  What I will discuss in this talk
is, specifically, how one may set about estimating the required expectation values in
perturbation theory.  For various reasons, perturbation theory is a rather limited tool for this
purpose, but I know of no better approximation scheme.  (The analysis given in
\cite{boyanovsky1998} avoids the use of perturbation theory by restricting itself to the
large-$N$ limit of O($N$)-symmetric scalar field theory, but does not appear to provide a useful
basis for a non-perturbative treatment of more general models;  the functional Schr\"odinger
approach described in \cite{samiullah1991} relies on a Gaussian anzatz for the wave functional
which seems very difficult to improve on in any systematic way and, as far as I know, has been
made to work only in a universe of 2 spatial dimensions.)  In setting about this enterprise, I
have encountered two significant problems.  The first is that the propagators of perturbation
theory as ordinarily conceived involve occupation numbers for quasiparticle modes which do not
evolve with time as they should when the system is driven away from thermal equilibrium.  The
second is that time evolution governed by a symmetrical Hamiltonian (here, $H(-\phi)=H(\phi)$)
from a symmetrical initial state (in which $\langle\phi\rangle$=0) cannot produce a nonzero
expectation value for $\phi$.  I will describe how I think these problems can be overcome, and
how a complete set of Feynman rules suitable for computing the time-dependent expectation values
can be obtained.

\section{Dissipative perturbation theory}

My calculations of non-equilibrium expectation values are based on the closed-time-path
formalism\cite{schwinger1961,keldysh1964,mathanthappa1962,chou1985,landsmann1987}. More
specifically, I use the path-integral technique described by Semenoff and
Weiss\cite{semenoff1985}, in which Green's functions are obtained from the generating
functional
\begin{equation}
Z(j_a)=\int[d\phi_a]\exp\left[i\bar{S}(\phi_a)+i\int d^4x\,j\cdot\phi\right]\,.
\label{generatingfunctional}
\end{equation}
Here, the single quantum field $\hat{\phi}(\bbox{x},t)$ is represented by three path
integration variables $\phi_a(\bbox{x},t)$ ($a=1, \cdots, 3$), which live on the Keldysh contour
in the complex time plane. The action $\bar{S}$ appearing in $Z(j_a)$ is
\begin{eqnarray}
\bar{S}(\phi_a)&=&\int d^3x\left[\int_0^{t_f}dt\,{\cal L}(\phi_1)-\int_0^{t_f}dt{\cal L}
(\phi_2)\right.\nonumber\\
&&\qquad\qquad\qquad+\left.i\int_0^{\beta}d\tau\,{\cal L}_E(\phi_3)\right]\,,
\end{eqnarray}
where ${\cal L}(\phi)$ is the original Lagrangian density (in our case, that given in
(\ref{lagrangian})), while ${\cal L}_E$ is the Euclidean version associated with the density
operator, namely
\begin{equation}
{\cal L}_E(\phi_3)=\frac{1}{2}\left(\partial_{\tau}\phi_3\right)^2+\frac{1}{2}\left(\nabla\phi
\right)^2-\frac{1}{2}m^2(0)\phi_3^2+\frac{\lambda}{4!}\phi_3^4\,.
\end{equation}
The time $t_f$ is the latest time that we wish to consider, and might well be taken to infinity.
I shall be particularly concerned with the real-time 2-point functions ($\alpha,\beta=1,2$)
given by
\begin{eqnarray}
G_{\alpha\beta}(x,x')&=&-\left.\frac{\partial}{\partial j_{\alpha}(x)}
\frac{\partial}{\partial j_{\beta}(x')}\ln Z(j_a)\right\vert_{j_a=0}\nonumber\\
&=&\pmatrix{
{\rm Tr}[\rho T(\hat{\phi}(x)\hat{\phi}(x'))]&{\rm Tr}[\rho\hat{\phi}(x')\hat{\phi}(x)]\cr
{\rm Tr}[\rho\hat{\phi}(x)\hat{\phi}(x')]&{\rm Tr}[\rho\bar{T}(\hat{\phi}(x)\hat{\phi}(x'))]}
\nonumber\\&&
\label{full2ptfunctions}
\end{eqnarray}
where $T$ and $\bar{T}$ denote respectively time-ordered and anti-time-ordered products of
the quantum field operator $\hat{\phi}(x)$ and $\rho$ is the initial density operator.  I shall
usually work with the spatial Fourier transforms of these functions, which can be written as
\begin{equation}
G_{\alpha\beta}(t,t';k)=H_{\beta}(t,t';k)\theta(t-t') + H_{\alpha}(t',t;k)\theta(t'-t)\,,
\end{equation}
in terms of a single complex function $H(t,t';k)$, with $H_1=H$ and $H_2=H^*$.  Other
expectation values can, of course, be obtained from appropriate derivatives of $Z(j_a)$.

The usual perturbative strategy of taking the quad\-ratic part of $\bar{S}(\phi_a)$ to define
the lowest-order approximation leads to propagators which can be expressed in the same way as
the full two-point functions,
\begin{equation}
g_{\alpha\beta}(t,t';k)=h_{\beta}(t,t';k)\theta(t-t') + h_{\alpha}(t',t;k)\theta(t'-t)\,.
\label{galphabeta}
\end{equation}
In this case, the function $h_k(t,t')$ is
\begin{equation}
h(t,t';k)=(1+n_k)f_k(t)f_k^*(t')+n_kf_k^*(t)f_k(t')\,,
\end{equation}
where $f_k(t)$ is a mode function satisfying
\begin{equation}
\left[\partial_t^2 +k^2-m^2(t)\right]f_k(t)=0
\end{equation}
and normalized by the Wronskian condition $\dot{f}(t)f^*(t)-f(t)\dot{f}^*(t)=-i$, while $n_k$
is the Bose-Einstein occupation number $n_k=\left(e^{\beta_0\omega_k}-1\right)^{-1}$ associated
with the initial state.  This is clearly unsatisfactory.  In the first place, since $m^2(t)$ is
positive and increasing with time, modes with $k^2<m^2(t)$ grow uncontrollably.  While such
behaviour might be expected during the course of the phase transition, it cannot be correct in
the initial high-temperature state.  Moreover, the occupation numbers are fixed at their
initial values, and cannot adequately describe the evolving state, except when $t$ is much
smaller than some characteristic relaxation time.  This $g_{\alpha\beta}$ is therefore not a
good approximation to the full two-point function.

To improve matters, we need mode functions and occupation numbers which, to a reasonable
approximation, describe the elementary quasiparticle excitations of the state which actually
exists at a given time.  To this end, I use a more sophisticated means of splitting the action
into an unperturbed part and an interaction \cite{lawrie1988,lawrie1989,lawrie1992}:
\begin{eqnarray}
\bar{S}(\phi_a)&=&\bar{S}_0(\phi_a)+\bar{S}_{\rm int}(\phi_a)\\
\bar{S}_0(\phi_a)&=&\bar{S}^{(2)}(\phi_a)\nonumber\\
&&+\frac{1}{2}\int\,dt\int\frac{d^3k}{(2\pi)^3}
\,\phi_{\alpha}(k){\cal M}_{\alpha\beta}(k)\phi_{\beta}(k)\\
\bar{S}_{\rm int}&=&\bar{S}^{(4)}(\phi_a)\nonumber\\
&&-\frac{1}{2}\int\,dt\int\frac{d^3k}{(2\pi)^3}
\,\phi_{\alpha}(k){\cal M}_{\alpha\beta}(k)\phi_{\beta}(k)\,,
\end{eqnarray}
where $\bar{S}^{(2)}$ and $\bar{S}^{(4)}$ are respectively the quadratic and quartic parts of
$\bar{S}$.  The counterterm ${\cal M}_{\alpha\beta}$ is to be chosen, according to an appropriate
renormalization prescription, to optimize $g_{\alpha\beta}$ as an approximation to
$G_{\alpha\beta}$.  The new unperturbed action can be expressed as
$\bar{S}_0=-{1\over 2}\int\,dt\,dk \phi_a{\cal D}_{ab}\phi_b$, where ${\cal D}$ is a differential
operator, and choosing ${\cal M}$ is clearly equivalent to choosing ${\cal D}$.  The real-time
propagators $g_{\alpha\beta}$ are solutions of
${\cal D}_{\alpha\gamma}(t,k)g_{\gamma\beta}(t,t';k)=-i\delta(t-t')$, and it is not hard to show
that the most general ${\cal D}$ which admits a solution of the form (\ref{galphabeta}) is
\begin{equation}
{\cal D}_{\alpha\beta}=\pmatrix{\partial_t^2+\beta_k-i\alpha_k&\gamma_k\partial_t
+\textstyle{\frac{1}{2}}\dot{\gamma}_k+i\alpha_k\cr
-\gamma_k\partial_t-\textstyle{\frac{1}{2}}\dot{\gamma}_k+i\alpha_k&
-\partial_t^2-\beta_k-i\alpha_k\cr}\,,
\label{diffop}
\end{equation}
where $\alpha_k(t)$, $\beta_k(t)$ and $\gamma_k(t)$ are undetermined, real functions.  These
functions can be determined self-consistently by using the counterterm ${\cal M}$ in
$\bar{S}_{\rm int}$ to cancel some part of the loop corrections to the full two-point functions
$G_{\alpha\beta}$. Details of the prescription used to do this are to some extent a matter of
choice, and the choices we make determine the extent to which properties of $G_{\alpha\beta}$
are reflected in $g_{\alpha\beta}$.  For simplicity, I choose $\beta_k(t)$ to have the form
$\beta_k(t)=\mu^2(t)+k^2$, so that $\mu(t)$ will be an effective quasiparticle mass, but in
principle a more accurate representation of the quasiparticle dispersion relation is possible.
The structure of the unperturbed action $\bar{S}_0$ obtained in this way is analogous to that
of the effective action obtained, for example, in \cite{hu1994} by integrating out extra
environmental degrees of freedom.

The propagator is now of the form (\ref{galphabeta}), with
\begin{eqnarray}
h&&(t,t';k)={\textstyle{1\over 2}}
\exp\left(-{\textstyle{{1\over 2}\int_{t'}^tdt''\gamma_k(t'')}}\right)\nonumber\\
&&\times\left[\left(N_k(t')+1\right)f_k(t)f_k^*(t')
+\left(N_k^*(t')-1\right)f_k^*(t)f_k(t')\right]\,.\nonumber\\
\label{littleh}
\end{eqnarray}
The mode function $f_k(t)$ is a complex solution of
\begin{equation}
\left[\partial_t^2+\mu^2(t)-{\textstyle{1\over 4}}\gamma_k^2(t)+k^2\right]f_k(t)=0\,,
\label{modeequation}
\end{equation}
while the new function $N_k(t)$ satisfies
\begin{eqnarray}
\left[\partial_t+2i\Omega_k(t)-\frac{\dot{\Omega}_k(t)}{\Omega_k(t)}+\gamma_k(t)\right]
\left[\partial_t\right.&+&\left.\gamma_k(t)\right]N_k(t)\nonumber\\
&&=2i\alpha_k(t)\,,
\label{nequation}
\end{eqnarray}
where $\Omega_k(t)=1/2f_k(t)f_k^*(t)$.  Evidently, the function $\gamma_k(t)$ plays the role of
a quasiparticle damping rate.  When the evolution with time is sufficiently slow, and the
coupling is sufficiently weak, it is possible to identify a time-dependent occupation number
$n_k(t)={1\over 2}\left[N_k(t)-1\right]$, and then (\ref{nequation}) can be cast approximately
in the standard form of a Boltzmann equation \cite{lawrie1989}.   The dissipative
formalism outlined here can be extended to complex scalar fields \cite{lawrie1997} and (with
somewhat greater difficulty) to spin-${1\over 2}$ fermions \cite{mckernan1998}.

It will soon be useful to know that, associated with ${\cal D}_{\alpha\beta}$ is an operator
\begin{equation}
\tensor{d}_{\alpha\beta}(t,k)=\pmatrix{\tensor{\partial}_t&\gamma_k(t)
\cr-\gamma_k(t)&-\tensor{\partial}_t\cr}\,,
\label{dlr}
\end{equation}
which has the property
\begin{eqnarray}
g(t,t'';k)\tensor{d}(t'',k)&&g(t'',t';k)\nonumber\\
&&=\left\{\matrix{
-ig(t,t';k)\,, & t>t''>t'\cr ig(t,t';k)\,, &  t'>t''>t\cr 0\,, & {\rm otherwise}\,.}
\right.\label{dleftright}
\end{eqnarray}

\section{Spontaneously unbroken symmetry}

As I observed earlier, a state which is initially symmetric under the operation $\phi\to-\phi$
necessarily evolves into another state with the same symmetry if time evolution is governed by
a Hamiltonian with the same symmetry.  Nevertheless, as the universe expands and cools, we
expect to encounter states in which the phenomena conventionally associated with spontaneously
broken symmetry (that is, with a non-zero value of $\langle\phi\rangle$) are realised, and I
shall refer to such states as having ``spontaneously unbroken symmetry''.  Here, I review briefly
the means I have found of treating a vacuum state of this kind in Minkowski spacetime
\cite{lawrie1988b}.  How such a state may be seen to emerge from a phase transition is the
central question that I shall address later on.

Intuitively, we can imagine a probability density for the field $\phi(\bbox{x})$ at each
spatial point $\bbox{x}$, which has symmetrically placed peaks both at
$\phi=\sigma/\sqrt{\lambda}$ and at $\phi=-\sigma/\sqrt{\lambda}$, where
$\sigma/\sqrt{\lambda}$ (with $\sigma$ of order 1) is the expectation value normally assigned to
$\phi$ in the broken-symmetry vacuum.  In a large spatial volume, interference between the two
components of the wave functional is negligible, and the doubly-peaked state (with spontaneously
unbroken symmetry) should be indistinguishable from the conventional broken-symmetry state in
which only one of the peaks is present.

More formally, we need a version of perturbation theory in which the unperturbed Lagrangian
describes free-particle excitations of the symmetrical vacuum, rather than the asymmetrical one.
To this end, I take the state of spontaneously unbroken symmetry to be characterised by the
expectation value of $\phi^2$, which is unconstrained by symmetry.  Instead of the conventional
field variable $\psi(x)$, defined by
\begin{equation}
\phi(x)=\sigma + \psi(x)\,,
\end{equation}
I deal with a field $\zeta(x)$ defined by
\begin{equation}
\phi^2(x)=U^2+2U\zeta(x)\,,
\label{zetadef}
\end{equation}
where $U^2=\langle\phi^2(x)\rangle$.  Taking $m^2(t)=m_0^2$ in (\ref{lagrangian}), I find
\begin{eqnarray}
{\cal L}=\frac{1}{2}&&\left(1+2U^{-1}\zeta\right)^{-1}\partial_{\mu}\zeta\partial^{\mu}\zeta
-\frac{\lambda}{6}U^2\zeta^2\nonumber\\
&& +U\left(m_0^2-\frac{\lambda}{6}U^2\right)\zeta
+\frac{i}{2}\delta^4(0)\ln\left(1+2U^{-1}\zeta\right)\,,\nonumber\\
\label{zetalagrangian}
\end{eqnarray}
where the last term comes from the functional Jacobian of the transformation and an irrelevant
constant has been dropped.  Since $\langle\zeta\rangle=0$, the linear term must vanish to
leading order, so we identify $U=\sqrt{6m_0^2/\lambda}(1+O(\lambda))$. The expansion of
(\ref{zetalagrangian}) in powers of $\lambda$ is
\begin{equation}
{\cal L}=\frac{1}{2}\partial_{\mu}\zeta\partial^{\mu}\zeta - \frac{1}{2}(2m_0^2)\zeta^2+\cdots\,.
\end{equation}
and we see at leading order that particles created by $\zeta$ from the spontaneously unbroken
vacuum have the same mass $\sqrt{2}m_0$ as those created from the broken-symmetry vacuum by
$\psi$.  It is not hard to convince oneself \cite{lawrie1988b} that these two types of
particle are completely indistinguishable:  they have the same mass (located by the pole of the
propagator) and the same $S$-matrix elements to all orders of perturbation theory.  In this
formulation, interactions arise from the kinetic term, and there are an infinite number of
vertices (though only a finite number of these appear at a given order in $\lambda$). The
analysis of spontaneously unbroken symmetry outlined here can be extended to the Higgs sector
of a spontaneously broken gauge theory \cite{lawrie1991}.  In accordance with a well-known
theorem of S. Elitzur \cite{elitzur1975}, non-zero expectation values need be assigned only to
operators which are invariant under the gauge and global symmetries of the theory. 

\section{Feynman rules for the phase transition}

As the universe cools, it seems intuitively clear that the probability density for $\phi$ starts
off with a single peak at $\phi=0$, and subsequently, say at a time $t_0$, develops the
double-peaked form I described earlier.  This picture is no more that an intuitive guide, because
I know of no way of actually computing this probability density in perturbation theory. Indeed,
the notion that it simply has either one or two symmetrically placed peaks may be too naive.
(It would certainly be too naive to suppose that the probability density for values of $\phi$
gives an adequate characterization of the total state:  for that one would need the probability
density for {\it functions} $\phi(\bbox{x})$ over an entire spatial section of the universe, which
is much more complicated than the one I have in mind.)  Specifically, the idea I wish to develop
is that at times earlier than $t_0$ (whenever this might be), the most probable value of $\phi$
is zero, and the appropriate way of attempting to use perturbation theory is by using the
Feynman rules based directly on the original Lagrangian (\ref{lagrangian}) in terms of $\phi$
itself, while at later times, the most probable values of $\phi$ are $\pm\sigma/\sqrt{\lambda}$
and the most appropriate perturbation expansion is that based on the formulation in terms of
$\zeta$.  Bringing this idea to fruition is awkward for two reasons.  One is that the
expectation value of $\phi^2$ is of order 1 in the expansion based on $\phi$, while it is of
order $\lambda^{-1}$ in the expansion based on $\zeta$ (or, for that matter, the expansion
based on $\psi$, if there were genuine symmetry breaking).  Thus, although both expansions are
in powers of $\lambda$, they are actually different expansions, corresponding to perturbations
about different lowest-order states, and one cannot expect to obtain a completely smooth
matching of these two approximations.  My strategy for dealing with this is the following.  To
one-loop order, at least, it is possible to identify by eye in the $\zeta$ formulation a
quantity $\sigma$ which it seems possible to interpret in the way I have described.  It should
also be possible (though I have actually done this only to lowest order) to derive an equation
of motion for $\sigma$, roughly of the form
\begin{equation}
\ddot{\sigma}(t)=f_0(\sigma)+\lambda f_1(\sigma)+\lambda^2f_2(\sigma)+\cdots\ ,
\label{sigmaequation}
\end{equation}
which is a systematic expansion in powers of $\lambda$, treating $\sigma$ as being of order 1
(though the right hand side is not a simple function of $\sigma(t)$, but rather a functional
which depends on the values of $\sigma$ at all times between $t_0$ and $t$).  The expectation
values we want to calculate can be similarly expressed.  The initial values of $\sigma$ and
$\dot{\sigma}$ at $t_0$ must, however, be obtained from the values of $\langle\phi^2\rangle$
and $d\langle\phi^2\rangle /dt$ at $t_0$ as calculated within the expansion based on $\phi$, and
are of order $\sqrt{\lambda}$.  With such initial conditions, there is certainly a solution to
(\ref{sigmaequation}), but this solution cannot be expressed as a power series in $\lambda$.
In any case, only a numerical solution is likely to be feasible.  My proposed strategy, then, is
to truncate both (\ref{sigmaequation}) and the expressions for the initial data at some chosen
(in practice, no doubt, low) order, and solve the resulting problem numerically.

The second source of difficulty is that the calculations using $\zeta$ for times after $t_0$
cannot be carried out in isolation from those using $\phi$ before $t_0$, because the
nonequilibrium density matrix {\it at} $t_0$ is not directly known.  It is therefore necessary
to treat the whole problem, making the change of variable (\ref{zetadef}) only on the real-time
part of the closed time path after $t_0$.  Thus, I need to evaluate a path integral of the
form
\begin{eqnarray}
Z(j_a,l_{\alpha})=\int\,&&[d\phi]\exp\left[i\int_0^{t_0}dt\left(\bar{\cal L}(\phi)
+j_a\phi_a\right)\right]\nonumber\\
&&\times\int\,[d\zeta]\times\exp\left[i\int_{t_0}^{t_f}dt\left(\bar{\cal L}(\zeta)
+l_{\alpha}\phi_{\alpha}\right)\right]\,.\nonumber\\&&
\end{eqnarray}
This, of course, is not simply the product of two independent path integrals, because of the
condition $\phi^2(t_0)=U^2(t_0)+2U(t_0)\zeta(t_0)$.  Evaluating it is no easy matter, and I have
not been able to do it as rigorously as I would like, but what I find is the following.  First,
the interactions can be extracted in the usual way, in terms of derivatives with respect to the
sources:
\begin{eqnarray}
Z(j_a,l_{\alpha})&=&\exp\left[i\bar{S}_{\rm int}^{\phi}\left(-i\frac{\delta}{\delta j}\right)
\right]\exp\left[i\bar{S}_{\rm int}^{\zeta}\left(-i\frac{\delta}{\delta l}\right)\right]
\nonumber\\&&\qquad\times Z_0(j_a,l_{\alpha})\,.
\end{eqnarray}
The exponential operators of course supply the vertices of Feynman diagrams, while the remaining
path integral $Z_0(j_a,l_{\alpha})$ supplies the propagators.  After doing the $\zeta$ integral,
I find
\begin{eqnarray}
Z_0&=&\exp\left[-\frac{1}{2}\int_{t_0}^{t_f}dt\,dt'l(t)g^{\zeta}(t,t')l(t')\right]\nonumber\\
&&\ \times\int\,[d\phi]\exp\left[i\int_0^{t_0}\left(\bar{\cal L}_0(\phi)+j_a\phi_a
+\phi^2_{\alpha}L_{\alpha}(t)\vphantom{\frac{1}{1}}\right)\right],\nonumber\\&&
\end{eqnarray}
where $L(t)$ is a distribution, concentrated at $t=t_0$, such that
\begin{eqnarray}
\int_0^{t_0}dt&&\,\phi^2_{\alpha}(t)L_{\alpha}(t)\nonumber\\
&&=\left.-\frac{i}{2}\int_{t_0}^{t_f}dt'\left(\frac{\phi_{\alpha}^2(t)}{U(t)}\right)
\tensor{d}_{\alpha\beta}(t)g^{\zeta}_{\beta\gamma}l_{\gamma}(t)\right\vert_{t=t_0}.
\end{eqnarray}
The action for $\phi$ clearly involves a differential operator ${\cal D}_{ab}(l)$ depending on
the source for $\zeta$.  Apart from some details which I will ignore, the result of doing the
final integration is (in an abbreviated notation which I hope is obvious)
\begin{equation}
Z_0=\exp\left[-\half lg^{\zeta}l-\half jg^{\phi}(l)j-\half {\rm Tr}\,\ln\left({\cal D}^{\phi}
(l)\right)\right]\,.
\end{equation}
Expanding $g^{\phi}(t,t';l)$ in powers of $l$, one obtains an infinite set of propagators.  The
first is the standard propagator for $\phi$, say $g^{\phi}(t,t')$.  The next is a composite
object, consisting of a vertex (involving $\tensor{d}_{\alpha\beta}(t_0)$) located at $t_0$, from
which two $\phi$ propagators and one $\zeta$ propagator emerge, and the remainder contain
strings of such vertices.  Similarly, expanding ${\rm Tr}\ln {\cal D}(l)$ yields a second set
of composite propagators, consisting of single loops of $\phi$ propagators with $\zeta$
propagators emerging, again from vertices located at $t_0$.  The connection between times before
$t_0$ and times after $t_0$ thus arises from these composite propagators rather than from
vertices.  Although the propagators are infinite in number, only as many of them appear in a
diagram of order $\lambda^n$ as can connect the vertices which yield this power of $\lambda$.

The interactions contained in $S_{\rm int}^{\zeta}$ naturally consist of an infinite set of
vertices, just as in the Minkowski-space theory.  Now, however, the quantity
$U^2(t)=\langle\phi^2(\bbox{x},t)\rangle$ depends on time.  It is still independent of
$\bbox{x}$, owing to the translational invariance of the Lagrangian and of the initial state.
Defining $v(t)=\lambda^{1/2}U(t)$,
the closed-time-path Lagrangian for $\zeta$ is
\begin{eqnarray}
\bar{\cal L}^{\zeta}&=&-\half \zeta_{\alpha}{\cal D}^{\zeta}_{\alpha\beta}\zeta_{\beta}
\nonumber\\
&&-\lambda^{-1/2}\left(\ddot{v}-m^2(t)v+{\textstyle{1\over 6}}v^3\right)\left(\zeta_1-\zeta_2
\right)\nonumber\\
&&+\half v^{-1}\left(\ddot{v}+M^2(t)v+{\textstyle{1\over 3}}v^3\right)\left(\zeta_1^2-\zeta_2^2
\right)+\cdots\ ,
\end{eqnarray}
where $M(t)$ is a renormalized mass for the $\zeta$ excitations.  The first term constitutes
$\bar{\cal L}^{\zeta}_0$, and yields the $\zeta$ propagator.  The other two terms that I have
shown explicitly are counterterms.  The one linear in $\zeta$ can be used to enforce the
condition $\langle\zeta\rangle=0$, which gives an equation of the form
$\ddot{v}-m^2(t)v+{1\over 6}v^3 = {\rm loops}$, to be solved for $v(t)$.  The term quadratic in
$\zeta$ can be used to cancel an appropriate part of the loop contributions to the $\zeta$
two-point function.  Normally, this would constitute a mass renormalization condition.  In this
formulation of the theory, however, the renormalized mass $M(t)$ actually appears in the linear
term.  Nevertheless, these two conditions do serve together to give an equation of motion for
$v$ and a relation between the renormalized mass $M(t)$ and the bare mass $m(t)$.

The equation of motion is somewhat complicated when written in terms of $v$, but simplifies
considerably when expressed in terms of the quantity
\begin{equation}
\sigma(t)=v(t)\left[1-\frac{\lambda}{2}\frac{I^{\zeta}_1(t)}{v^2(t)}+O(\lambda^2)\right]\,,
\end{equation}
where $I_1^{\zeta}(t)$ denotes the single-propagator bubble
\begin{equation}
I_1^{\zeta}(t)=\int\frac{d^3k}{(2\pi)^3}g^{\zeta}(t,t;k)\,.
\end{equation}
In the Minkowski-space theory, Green's functions involving the operator $\phi^2$ are rendered
ultraviolet finite by a suitable combination of additive and multiplicative renormalizations.
The same procedure works here (to one-loop order, at least) and leads to a renormalized
equation of motion which is
\begin{equation}
\ddot{\sigma}_R+M^2(t)\sigma_R-\frac{1}{3}\sigma_R^3+O(\lambda^2)=0\,.
\label{eom}
\end{equation}
The relation between $M(t)$ and $m(t)$ can also be renormalized, leading to a gap equation
\begin{eqnarray}
M^2(t)&=&\frac{1}{2}\sigma_R^2(t)+\frac{1}{2}\tilde{I}_1^{\zeta}(t)\nonumber\\
&&-\frac{1}{2}a^2(t)\hat{m}^2
\left\{1+\frac{\lambda}{16\pi^2}\left[c-\ln\left(a(t)\hat{m}\right)\right]\right\}
\nonumber\\&&+O(\lambda^2)\,,
\label{gap}
\end{eqnarray}
where $\hat{m}$ is the physical particle mass in the Minkowski-space vacuum (which locates the
pole of the $\zeta$ propagator) and $c=1+\ln 2-\sqrt{3}\pi/2$.  The integral
$\tilde{I}_1^{\zeta}(t)$ is a subtracted version of $I_1^{\zeta}(t)$, which is ultraviolet
finite, provided that the large-$k$ behaviour of the propagators is similar to that in the
equilibrium theory.
In the $\phi$-based perturbation theory before $t_0$, one can calculate directly
\begin{equation}
v^2(t)=I_1^{\phi}(t)+O(\lambda)
\end{equation}
and obtain a renormalized gap equation for the mass $\mu(t)$ of $\phi$ quasiparticles, which
reads
\begin{eqnarray}
\mu^2(t)&=&\frac{1}{2}\tilde{I}_1^{\phi}(t)-\frac{1}{2}a^2(t)\hat{m}^2
\left\{1+\frac{\lambda}{16\pi^2}\left[c-\ln\left(a(t)\hat{m}\right)\right]\right\}
\nonumber\\&&+O(\lambda^2)\,.
\label{mugap}
\end{eqnarray}
Initial conditions for the solution of (\ref{eom}) are obtained by requiring continuity of
$v^2(t)$ and its time derivative at $t_0$.  The differential equation to be solved for
$g^{\zeta}(t,t')$ also requires initial conditions. These can be found by noting that the
connected two-point function for $\phi^2$ can be expressed as
\begin{equation}
\langle\phi^2(\bbox{x},t)\phi^2(\bbox{x}',t')\rangle_c=
\langle\phi^2(\bbox{x},t)\zeta(\bbox{x}',t')\rangle_c\cdot 2\lambda^{-1/2}v(t')\,,
\end{equation}
and so on, depending on whether its time arguments are greater or smaller than $t_0$.  Initial
conditions for $g^{\zeta}(t,t')$ are then obtained by requiring this two-point function and its
first derivatives to be continuous at $t_0$.

In summary, the evaluation of an expectation value in the $\phi$-based perturbation theory for
early times works as follows.  We first calculate the Feynman diagrams which appear at the
desired order in $\lambda$, obtaining a formal expression in terms of the propagators
$g_{\alpha\beta}(t,t';k)$.  In order to convert this formal expression into a concrete one, we
need values for five quantities, namely the mode functions $f_k(t)$, the generalised occupation
numbers $N_k(t)$, the quasiparticle mass $\mu(t)$ and damping rate $\gamma_k(t)$ and the
auxiliary functions $\alpha_k(t)$. These values are found by solving self-consistently a set
of five equations.  They are: (i) the equation (\ref{modeequation}) for the mode functions;
(ii) the equation (\ref{nequation}) for the $N_k(t)$; (iii) the gap equation (\ref{mugap})
for the quasiparticle mass; (iv) a constraint equation for $\gamma_k(t)$ and (v) a constraint
equation for $\alpha_k(t)$.  Equations (iv) and (v) are somewhat complicated, so I have not
given them explicitly (but details can be found in \cite{lawrie1989}).  They are of the same
kind as the gap equation (\ref{mugap}), since all three equations amount to the requirement
that $\mu(t)$, $\alpha_k(t)$ and $\gamma_k(t)$ approximate appropriate parts of the full
self-energy matrix.  The equations for $\alpha_k(t)$ and $\gamma_k(t)$ receive contributions
from the absorptive parts of the self-energy (which are of at least two-loop order) in place
of the one-loop integral $\tilde{I}_1^{\phi}$ that appears in (\ref{mugap}).  Of course, these
integrals involve the very quantities we need to find so, although the solutions are fully
determined, they cannot be obtained in closed form (except, perhaps, in very special
circumstances).  In principle, however, the equations are susceptible of numerical solution.

Just how difficult this numerical problem is naturally depends on what further approximations
one is willing to make.  To give an idea of what is involved, let me consider the limit of slow
evolution and weak coupling.  In that case, I can define occupation numbers $n_k(t)$ by
\begin{equation}
N_k(t)\approx\frac{2n_k(t)+1}{1-i\gamma_k(t)/2\sqrt{k^2+\mu^2(t)}}
\end{equation}
and (\ref{nequation}) reduces to
\begin{equation}
\frac{\partial n}{\partial t}\approx\frac{1}{2}\left[\frac{\alpha_k(t)}{\sqrt{k^2+\mu^2(t)}}
-\gamma_k(t)\right]-\gamma_k(t)n_k(t)\,.
\label{approxnequation}
\end{equation}
In turn, $\alpha_k(t)$ and $\gamma_k(t)$ are given by scattering integrals which involve
$n_k(t)$, such that (\ref{approxnequation}) takes the form of a Boltzmann equation.  At this
point, simultaneous numerical solution of (\ref{modeequation}), (\ref{mugap}) and
(\ref{approxnequation}) yields estimates of $n_k(t)$ and $\mu(t)$, from which $\alpha_k(t)$ and
$\gamma_k(t)$ can be reconstructed.  The possibility of expressing these two functions
explicitly in terms of the $n_k(t)$ results from the fact that they are of order $\lambda^2$ and
can reasonably be set to zero within scattering integrals that are already multiplied by
$\lambda^2$.  Similarly, $\gamma_k^2(t)$ might be omitted from (\ref{modeequation}).  If one is
not satisfied with approximations of this sort, then all five equations must be numerically
evolved together.  Two of the equations we have to solve, namely (\ref{modeequation}) and
(\ref{nequation}) are differential equations which need initial conditions.  These, of course,
are obtained from the initial equilibrium state and the considerations involved are discussed
in detail in \cite{lawrie1992}.

Expectation values at later times are obtained from the $\zeta$-based Feynman rules in almost the
same way.  The differences are: (a) we use the gap equation (\ref{gap}) in place of
(\ref{mugap}); (b) these rules contain a sixth unknown function, $v(t)$ (or $\sigma(t)$),
and, correspondingly, there is a sixth equation, the equation of motion (\ref{eom}), to be
solved simultaneously with the other five.  Initial conditions for this set of equations are
obtained from continuity requirements at $t=t_0$.

\section{Discussion}

I have explained in outline how I think that perturbation theory can be used to describe the
nonequilibrium dynamics of a symmetry-breaking phase transition in an expanding universe.  A
straightforward application of standard perturbative procedures is inadequate for two
reasons.  One is that low-order calculations (which are all that one can contemplate
tackling) will reflect the evolving state of the system only if the propagators are defined
self-consistently in terms of quasiparticle modes which, to a reasonable approximation,
represent the elementary excitations of the actual state.  The other is that spontaneous
symmetry breaking as conventionally understood cannot come about if the system starts from
a symmetrical state and its evolution is governed by a symmetrical Hamiltonian.  The Feynman
rules I have outlined are designed to overcome both of these difficulties.

There are, however, other problems inherent in any attempt to treat this problem perturbatively,
that I do not think can altogether be overcome.  One is that the perturbative expansions
appropriate to the initial symmetrical state and to the emerging state that I describe as having
spontaneously unbroken symmetry are really two different approximations.  This leads to some
formal difficulties, beyond the ones I have mentioned, that I do not propose to discuss here
(though I do discuss them in a more detailed forthcoming paper).

A second problem is apparent from equation (\ref{gap}), where we see a negative contribution to
the quasiparticle mass $M^2(t)$, whose magnitude increases with time.  The mode functions from
which the propagators are constructed obey equations of the form (\ref{modeequation}) containing
a squared mass which will be negative for some period of time around $t_0$, and will grow,
potentially to very large values.  There is clearly the danger that perturbation theory will
become meaningless if this growth continues for too long.  The growth cannot continue
indefinitely, because it causes the positive contributions to (\ref{gap}) to increase in
magnitude faster than the negative ones. It is therefore not inevitable that perturbation theory
need become completely unreliable, but concrete numerical calculations (which I have not yet
undertaken) would be needed to assess the extent of the danger.
Because of this difficulty, Boyanovsky {\it et. al.} \cite{boyanovsky1998} have invested
considerable effort in obtaining numerical solutions for the case of the large-$N$ limit of the
${\rm O}(N)$-symmetric theory. The equations they solve (which are exact in the large-$N$ limit)
include ones which are rather similar to (\ref{eom}) and (\ref{gap}), though their meanings are
slightly different.  In particular, $\sigma$ corresponds in that case simply to the expectation
value of one component of $\phi$, which is identically zero for a symmetric initial state.  In
my language, therefore, their calculation is equivalent to working only with the $\phi$
formulation and this, of course, is possible only with an exactly soluble model.  The initial
conditions adopted by these authors correspond essentially to a vacuum state, with
$h(t,t';k)=\half f_k(t)f_k^*(t')$ in place of (\ref{littleh}).  While initial conditions
appropriate to an initial high-temperature state could no doubt also be treated by their
methods, the corresponding occupation number would not be seen to evolve with time, because the
large-$N$ theory contains no scattering mechanism for the redistribution of energy.  For this
reason, the final state in this calculation is one that oscillates perpetually: the
thermalization that one would expect in a more realistic model does not occur.

Despite the difficulties I have alluded to, the analysis I have summarized does lead, at the
lowest nontrivial order, to a well defined set of equations which might be solved numerically and
should provide a description of the quantum dynamics of a phase transition with at least the
qualitative features (including thermalization of the final state) that one would expect.  In
principle (though perhaps with some technical difficulty) the approximation scheme can be
extended to higher orders, and it should also be possible to extend this scheme to more realistic
models of particle physics.  The inherent deficiencies of perturbation notwithstanding, I know
of no other approach which has this potential.

\end{document}